\documentclass[aps,prl,twocolumn,superscriptaddress]{revtex4} 
\usepackage[T1]{fontenc}
\usepackage{amssymb}

\usepackage{hyperref}
\hypersetup{
    bookmarks=true,         
    unicode=true,          
    pdftoolbar=false,        
    pdfmenubar=true,        
    pdffitwindow=true,      
    pdftitle={My title},    
    pdfauthor={Author},     
    pdfsubject={Subject},   
    pdfnewwindow=true,      
    pdfkeywords={keywords}, 
    colorlinks=true,       
    linkcolor=red,          
    citecolor=black,        
    filecolor=magenta,      
    urlcolor=cyan           
}

\usepackage[all]{hypcap}
\usepackage{graphicx, subfigure}
\usepackage{amsmath}
\usepackage{natbib}
\usepackage{layouts}
\usepackage{threeparttable}
\usepackage{footnote}
\usepackage{booktabs}
\usepackage{scrextend}
\usepackage{hyperref}

\begin{document}

\title{First-principles study of configurational disorder in B$_{4}$C\\ using a superatom-special quasirandom structure method}

\author{A. Ektarawong} 
\email{anekt@ifm.liu.se}
\affiliation{Thin Film Physics Division, Department of Physics, Chemistry and Biology (IFM), Link\"oping University, SE-581 83  Link\"oping, Sweden}
\author{S. I. Simak}
\affiliation{Theoretical Physics Division, Department of Physics, Chemistry and Biology (IFM), Link\"oping University, SE-581 83  Link\"oping, Sweden}
\author{L. Hultman}
\affiliation{Thin Film Physics Division, Department of Physics, Chemistry and Biology (IFM), Link\"oping University, SE-581 83  Link\"oping, Sweden}
\author{J. Birch}
\affiliation{Thin Film Physics Division, Department of Physics, Chemistry and Biology (IFM), Link\"oping University, SE-581 83  Link\"oping, Sweden}
\author{B. Alling}
\affiliation{Thin Film Physics Division, Department of Physics, Chemistry and Biology (IFM), Link\"oping University, SE-581 83 Link\"oping, Sweden}


\begin{abstract} 
\indent{Configurationally disordered crystalline boron carbide, with the composition B$_{4}$C, is studied using first-principles calculations. We investigate both dilute and high concentrations of carbon-boron substitutional defects. For the latter purpose, we suggest a superatom's picture of the complex structure and combine it with a special quasi-random structure approach for disorder. In this way, we model a random distribution of high concentrations of the identified low-energy defects: 1) Bipolar defects and 2) Rotation of icosahedral carbon among the three polar-up sites. Additionally, the substitutional disorder of the icosahedral carbon at all six polar sites, as previously discussed in the literature, is also considered. Two configurational phase transitions from the ordered to the disordered configurations are predicted to take place upon increasing temperature using a mean-field approximation for the entropy. The first transition, at 870 K, induces substitutional disorder of the icosahedral carbon atoms among the three polar-up sites, meanwhile the second transition, at 2325 K, reveals the random substitution of the icosahedral carbon atoms at all six polar sites coexisting with bipolar defects. Already the first transition removes the monoclinic distortion existing in the ordered ground state configuration and restore the rhombohedral system (\emph{R${3}$m}). The restoration of inversion symmetry yielding the full rhombohedral symmetry (\emph{R$\bar{3}$m}) on average, corresponding to what is reported in the literature, is achieved after the second transition. Investigating the effects of high pressure on the configurational stability of the disordered B$_{4}$C phases reveals a tendency to stabilize the ordered ground state configuration as the configurationally ordering/disordering transition temperature increases with pressure exerted on B$_{4}$C. The electronic density of states, obtained from the disordered phases indicates a sensitivity of band gap to the degree of configurational disorder in B$_{4}$C.}

\end{abstract}

\maketitle 

\section{I. Introduction}
\indent{Boron carbide is a class of materials considered for several important applications. The high cross-section for thermal neutron reaction of the isotope $^{10}$B makes boron carbide relevant as a new generation of neutron detectors~\cite{Lacy2011, Carina2012} possibly replacing the present dominating $^{3}$He-based technologies suffering from the $^{3}$He-crisis. Even though only $^{10}$B gives the contribution to detect neutrons, B$_{4}$C is still considered as the most suitable candidate rather than pure $^{10}$B due to its excellent properties in terms of stability and non-toxicity~\cite{Knotek1997}. In the past few years, prototypes of B$_{4}$C-based neutron detectors in the form of multi-layer thin solid films have been proposed~\cite{Lacy2011, Carina2012}. B$_{4}$C is also used as a shielding material in nuclear reactors due to its ability to absorb neutrons and its unusual properties to heal itself from radiation damage~\cite{Simeone2000, Carrard1995, Emin2006}. Moreover, B$_{4}$C processes several outstanding properties such as high hardness, low density, high melting point, low wear coefficient, and high chemical stability. Such properties make it usable to, for example, a light-weight armour, a wear resistance material, a cutting tool material~\cite{Zorzi2005}, and a candidate material for high-temperature-electronic~\cite{Brand2001} and -thermoelectric~\cite{Werheit1995} devices.}\\
\begin{figure}[htb]
		\centering
		\includegraphics[width=\linewidth]{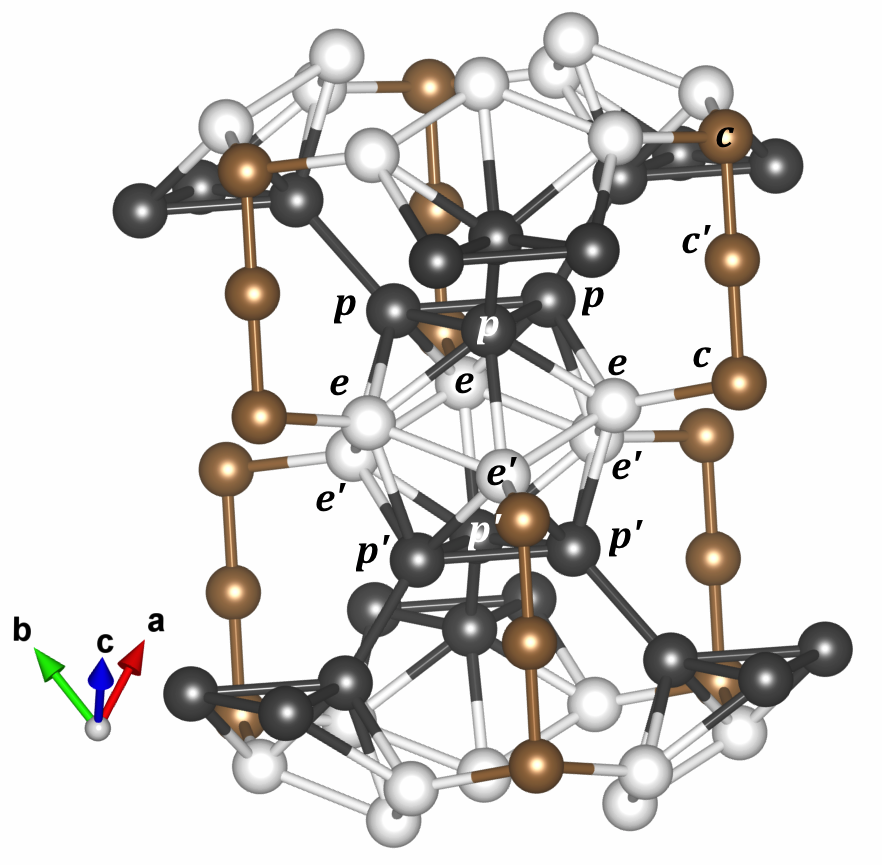}
		\caption{(Color online) The icosahedral structure of boron carbide. Black, white, and brown spheres represent the polar (-up \emph{p}, -down \emph{p'}), the equatorial (-up \emph{e}, -down \emph{e'}) and the chain (-end \emph{c}, -centre \emph{c'}) sites, repectively.}
		\label{fig:1}
	\end{figure}
\indent{To improve the performance of the B$_{4}$C-materials in such applications, a detailed understanding of the structure on the atomic level is needed. The structures of B$_{4}$C and related compositions of boron carbide, e.g. B$_{13}$C$_{2}$ have been studied both experimentally and theoretically~\cite{Bylander1990, Bylander1991, Lazzari1999, Mauri2001, Vast2009, Shirai_12010, Shirai_22010, Domnich2011, Decarlos2012} but the substitutional disorder of the carbon atoms has not been sufficiently investigated. Revealed by diffraction measurements~\cite{Clark1943, Yakel1975, Morosin1986, Morosin1995}, B$_{4}$C has a rhombohedral symmetry with a suggested space group \emph{R$\bar{3}$m}, \#166 in International Tables for Crystallography. There are 15 atoms in the rhombohedron unit cell, i.e. a 12-atom icosahedra located at vertices of the rhombohedron with a 3-atom linear chain aligned along the [111] direction in the centre of the rhombohedron (see FIG.~\ref{fig:1}). Each icosahedron is formed by 2 types of crystallographic sites, i.e. 6 sites for each type, namely, polar and equatorial sites. Atoms in the icosahedron sitting on polar sites form inter-icosahedral bonds to polar atoms in the neighboring icosahedra. On the other hand, atoms sitting on the equatorial sites are linked to the 3-atom chains. An arrangement of boron and carbon atoms in the chain is specified, by X-ray diffraction technique~\cite{Morosin1987, Larson1986}, to be C-B-C. The third carbon atom, as a consequence, must be substituted in the icosahedron. Experimentally identifying an atomic position of the carbon atom in the icosahedron is a formidable task owing to very close atomic form factors between these two light neighboring elements. First-principles calculations of energetics of B$_{4}$C has been done by Bylander \emph{et al.}~\cite{Bylander1990}. They suggested that the substitution of the third carbon atom is at the polar site in the unit cell. The configuration, often denoted as (B$_{11}$C$^{p}$)+ (C-B-C), yields the lowest total energy. The same conclusion has been later confirmed by theoretical inspection of vibrational properties~\cite{Lazzari1999} and by a first-principles analysis of NMR spectra~\cite{Mauri2001}. Such a substitution of the carbon atom also reduces the symmetry, by a small distortion, from rhombohedral to base-centered monoclinic (space group \emph{Cm}, \#8). However, an experimental observation of that monoclinic B$_4$C phase has never been reported. The reason has been suggested~\cite{Lazzari1999} to be  due to configurational disorder, and in particular, a disordered substitution of a carbon atoms among all six different polar sites with equal concentration. Such disorder should restore the higher rhombohedral symmetry found in experiments. Apart from the substitutional disorder of the carbon atom among the polar sites in the icosahedra, substitution of the polar carbon from one icosahedron to a neighboring icosahedron forming (B$_{10}$C$^{p}_2$)+(B$_{12}$), so-called bipolar defects, are demonstrated to exist in the structure~\cite{Mauri2001}.}\\
	\begin{figure}[htb]
		\centering
		\includegraphics[width=0.75\linewidth]{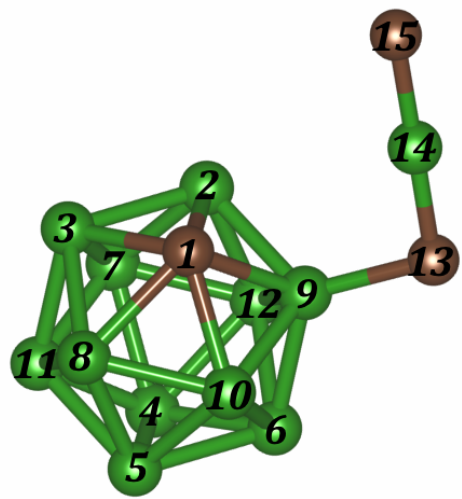}
		\caption{(Color online) A 12-atom icosahedron and a 3-atom chain made of boron (green) and carbon (brown) atoms. A set of numbers (1, 2, 3) denotes the polar-up sites. Similarly, (4, 5, 6) denote the polar-down sites. The equatorial-up (7, 8, 9), the equatorial-down (10, 11, 12), and the chain (13, 14, 15) sites are included in the figure corresponding to the number in the parenthesis. A dilute defect, namely "a rotation of the icosahedral carbon atom among the polar-up sites", takes place when a carbon atom at position1 swaps its position with a boron atom at position2 or 3.}
		\label{fig:2}
	\end{figure}
\indent{Disorder in boron carbide has been studied both experimentally~\cite{Werheit1992, Morosin1996, Werheit1999} and theoretically~\cite{Saal2007, Musiri2007, Vast2009, Vast2011, Widom2012}. Moreover, thin films of B$_{4}$C, which are more or less under-stoichiometric, are amorphous when grown at low temperature, i.e. topologically disordered~\cite{Camille2013}. Despite the evidence that existing samples of B$_{4}$C are configurationally disordered, no theoretical method has been suggested to directly calculate the properties of  such disordered structures. Therefore, we present in this work a theoretical technique that provides a path to investigate configurationally disordered B$_{4}$C, based on first-principles calculations. We start with a detailed study of dilute defects, with focus on the site displacement of the icosahedral carbon atom. Various types of dilute defects are introduced into sufficiently big supercells of ideally ordered B$_{4}$C. As seen by determining defect formation energies, we single out what kinds of defects that are most likely to appear in the structure during synthesis. We demonstrate that a rotation of the icosahedral carbon atom among the polar-up sites (position1, 2, and 3 in FIG.~\ref{fig:2}) and a bipolar defect (see FIG.~\ref{fig:3}) are the most likely ones. Note that in this work for all cases, the ratio of the boron to the carbon concentrations keeps the 4:1 stoichiometry (20\% atomic carbon concentration). Also, all of the configurations considered are electrically neutral. In the next step, to study substitutionally disordered configurations of B$_{4}$C, we suggest a method to distribute the identified most important defect types with high concentrations. The method, superatom-special quasi random structure (SA-SQS), is based on a combination of the special quasi-random structure (SQS) approach, originally suggested by Zunger \emph{et al.}~\cite{Zunger1990}, and a concept of superatoms of different icosahedral structures, which are identified through dilute defect calculations. This allows us to investigate theoretically, one step closer to the real situation, the properties, such as symmetry and electronic density of states, of disordered B$_{4}$C phases and to model order-disorder transitions.}
	\begin{figure}[htb]
		\centering
		\includegraphics[width=0.6\linewidth]{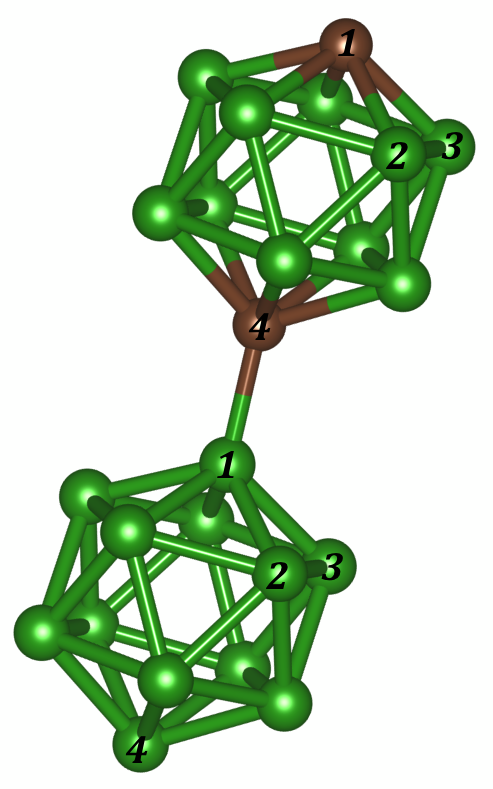}
		\caption{(Color online) A bipolar defect involved with two icosahedra, i.e. (B$_{11}$C$^{p}_2$)+(B$_{12}$). The labeled numbers and colors indicate the same positions and types of atoms, respectively, as described in FIG.~\ref{fig:2}. The 3-atom chains are not shown.}
		\label{fig:3}
	\end{figure}
\section{II.  Computational details}
	\indent{All calculations were done based on density functional theory. The projector augmented wave (PAW) method, as implemented in the Vienna \emph{ab initio} simulation package (VASP)~\cite{Blochl1994, Kresse1993}, was utilized to calculate the total energy of the boron carbide systems. The Perdew-Becke-Ernzerhof (PBE96) generalized gradient approximation (GGA)~\cite{Perdew1996} was chosen for the exchange-correlation functional. The monoclinic-distorted B$_{4}$C, based on a 15-atom unit cell [(B$_{11}$C$^{p}$)+(C-B-C)], was used as a starting point for the further calculations. The atomic coordinates and the cell shape were fully relaxed in each structural case. The equilibrium volume was then obtained from a curve plotted for the total energies versus the fixed-volumes. The energy cutoffs of 400 eV were set for plane wave calculations. The total energy was converged within an accuracy of 1 meV/atom, respecting to both the energy cutoff and the number of k-points used for the integration over the Brillouin zone. A 5x5x5 Monkhorst-Pack \textbf{k}-point mesh~\cite{Monkhorst1976} were used for the case of 15-atom unit cell, and of supercell sizes up to only 2x2x2 (120 atoms), meanwhile the supercell sizes larger than 2x2x2, i.e. 3x3x3 (405 atoms) and 4x3x3 (540 atoms), were sampled with a 3x3x3 Monkhorst-Pack \textbf{k}-point mesh. For electronic density of states calculations, the tetrahedron method for Brillouin zone integrations suggested by Bl\"ochl~\cite{PE1994} was used with a 9x9x9 Monkhorst-Pack \textbf{k}-point mesh for a 15-atom unit cell and up to 2x2x2 sizes of supercells, and with a 5x5x5 Monkhorst-Pack \textbf{k}-point mesh for the supercell sizes larger than 2x2x2. Additionally, the modified Becke-Johnson exchange potential, MBJ~\cite{Becke2006, Tran2009} in combination with the GGA-PBE correlation was applied to density of states calculations for all configurations with a 5x5x5 Monkhorst-Pack \textbf{k}-point mesh. For benchmarking, the hybrid functional HSE06~\cite{HSE03, HSE06} was applied to density of states calculation for the 15-atom ordered unit cell.}\\
	\indent{To study the effects on the B$_{4}$C system, when different kinds of a dilute defect were introduced, supercell sizes up to 2x2x2 unit cells (120 atoms) were mainly used in the calculations. This size of supercells ensures that the interactions between the defects are small and the defect formation energies were obtained with a size dependent uncertainty of less than 40 meV, as judged from comparison to 3x3x3 supercells (405 atoms) calculations. As mentioned in the previous section that for all cases the atomic carbon concentration is kept fixed at 20\%, each type of a dilute defect was thus created with respect to the site displacement of an icosahedral carbon atom in one specific icosahedron in the supercell. During the calculations, only the atomic coordinates were allowed to relax and the input volume was fixed based on the equilibrium one. The defect formation energy (\emph{$\Delta$E$_{defect}$}) was directly calculated by the equation;}\\
	\begin{equation}
		{\Delta}E_{defect}=E_{{defect}}-E_{{[B_{11}C^{p}]+[C-B-C]}},
		\label{eq:1}
	\end{equation}
where E$_{defect}$ and E$_{{[B_{11}C^{p}]+[C-B-C]}}$ are defined as the total energy of the system with and without a defect, respectively. 14 types, in total, of dilute defects were considered as listed in TABLE~\ref{tab:1}, including their formation energies. Regarding the stability at zero pressure and zero temperature conditions for all configurations of B$_{4}$C, the formation energy with respect to a decomposition into elemental phases, i.e. in this work, $\alpha$-boron and diamond, was also considered. Visualizations of various configurations were obtained with the VESTA package~\cite{Vesta2011}. 

\begin{figure*}[ht!]
     \begin{center}
        \subfigure[]{%
            \label{fig:4_1}
            \includegraphics[width=0.35\textwidth]{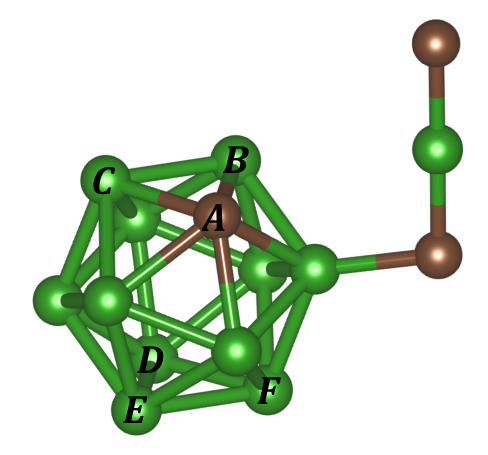}
        }%
        \subfigure[]{%
           \label{fig:4_2}
           \includegraphics[width=0.35\textwidth]{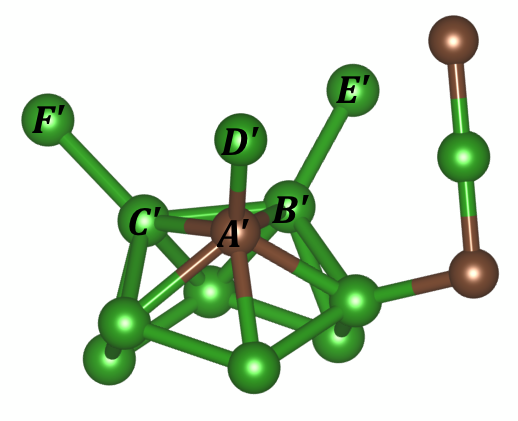}
        }\\ 
    \end{center}
    \caption{%
        (Color online) Two kinds of superatom bases for constructing the disordered configurations in B$_{4}$C: (a) Basis-1, and (b) Basis-2.  Green and brown spheres represent boron and carbon atoms, respectively. The types of superatom are distinguished by the respective position of the icosahedral carbon atom. The alphabet \emph{A} to \emph{F} (\emph{A'} to \emph{F'}) denotes the types of superatom and also the position of the icosahedral carbon atom for Basis-1 (Basis-2).
     }%
   \label{fig:4}
\end{figure*}

\begin{figure*}[ht!]
     \begin{center}

        \subfigure[]{%
            \label{fig:5_1}
            \includegraphics[width=0.3\textwidth]{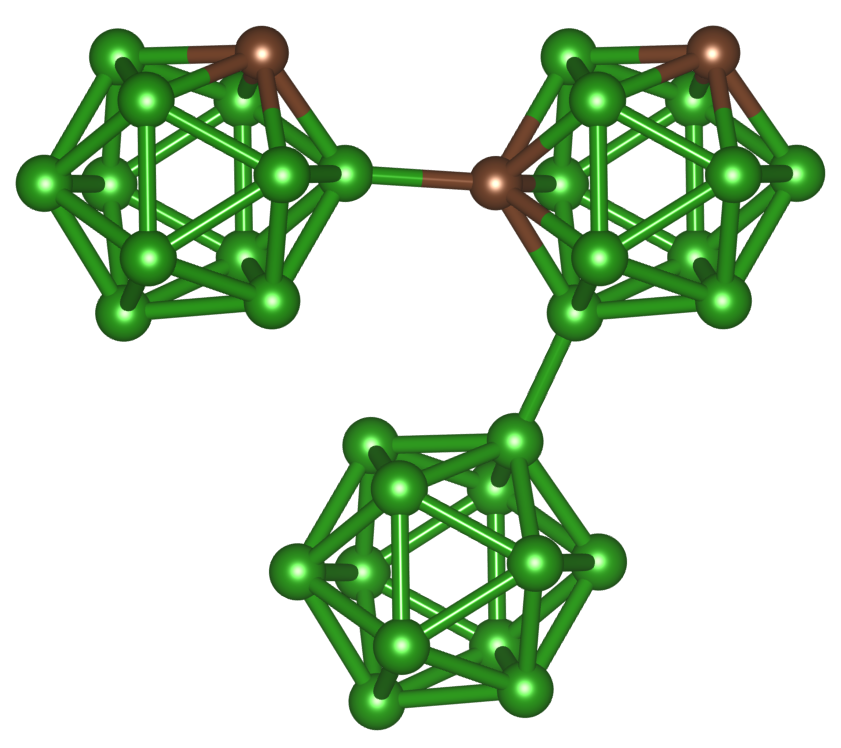}
        }%
        \subfigure[]{%
           \label{fig:5_2}
           \includegraphics[width=0.3\textwidth]{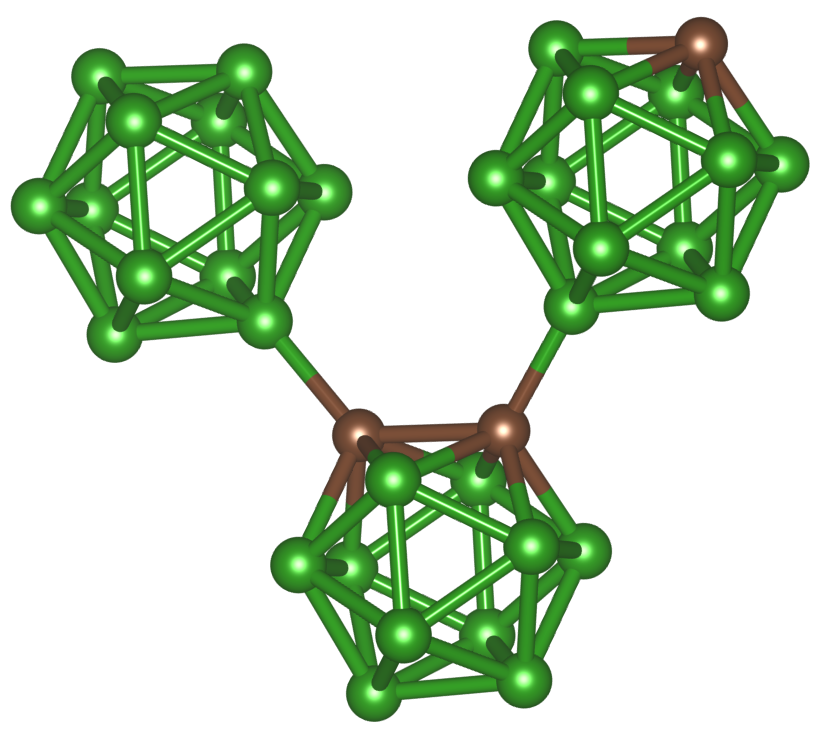}
        }
        \subfigure[]{%
           \label{fig:5_3}
           \includegraphics[width=0.3\textwidth]{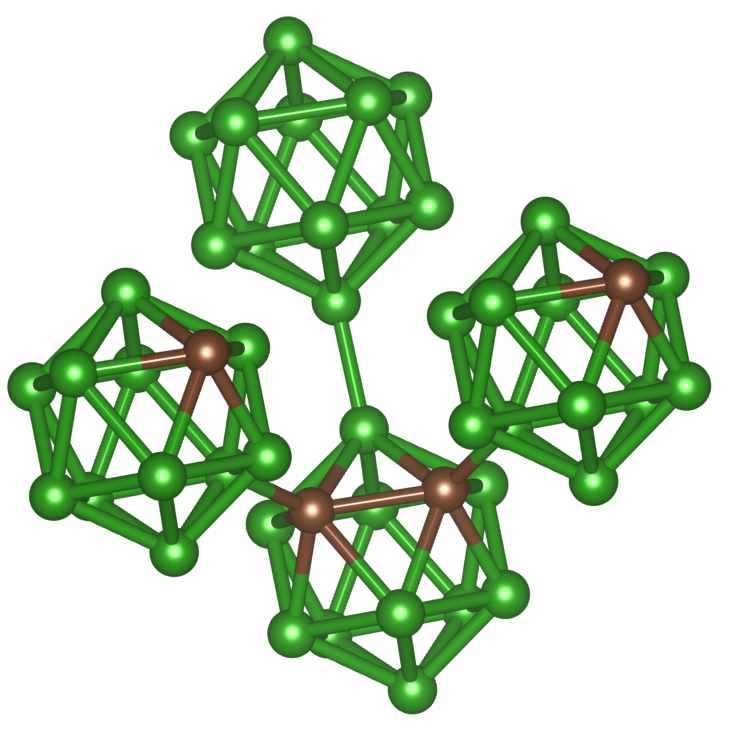}
        }\\ 
    \end{center}
    \caption{%
        (Color online) Three cases of the distorted bipolar defect: (a) Distorted Bipolar Defect \#1, (b) Distorted Bipolar Defect \#2, and (c) Distorted Bipolar Defect \#3. Green and brown spheres represent boron and carbon atoms, respectively. The 3-atom chains are not shown.
     }%
   \label{fig:5}
\end{figure*}
\begin{table}
\caption{Studied 14 types of a dilute defect in 2x2x2 supercells and their formation energies with respect to B$_{11}$C$^{p(1)}$+(C-B-C) structure, which is used as a reference. The superscripts \emph{p} and \emph{e} denote the \emph{polar} and the \emph{equatorial} sites in the icosahedron, respectively. The superscript of the number in the parentheses corresponds to the position in the icosahedron as indicated in FIG.~\ref{fig:2}. For three special cases: Distorted Bipolar Defects, see FIG.~\ref{fig:5}.}
\begin{minipage}{8cm}
\begin{tabular}{l @{\hspace{0.5cm}} c}
\hline
\hline
Defective structure & Formation energy (eV)\\
\hline
Non-defective structure & \\
B$_{11}$C$^{p(1)}$+(C-B-C) & 0 \\
\hline
Bipolar defect & \\
B$_{10}$C$_{2}$$^{p(1, 4)}$+B$_{12}$+(C-B-C)$_{2}$ & 0.229 \\
 & 0.262 \footnote{{This work - 3x3x3 supercell} \label{note:1}} \\
 & $\thicksim$0.23 \footnote{Ref.~\cite{Vast2011} - 2x2x2 supercell (LDA-PP) \label{note:2}} \\
 & 0.477 \footnote{Ref.~\cite{Widom2012} - 3x1x1 supercell (GGA-PW91) \label{note:3}} \\
 \hline
Polar sites & \\
B$_{11}$C$^{p(2, 3)}$+(C-B-C) & 0.232 \\
 & 0.248 \footref{note:1} \\
 & 0.135 \footref{note:3} \\
B$_{11}$C$^{p(4)}$+(C-B-C) & 1.217  \\
B$_{11}$C$^{p(5, 6)}$+(C-B-C) & 0.464  \\
\hline
Equatorial sites & \\
B$_{11}$C$^{e(7)}$+(C-B-C) & 0.674 \\
B$_{11}$C$^{e(8, 9)}$+(C-B-C) & 0.714 \\
B$_{11}$C$^{e(10)}$+(C-B-C) & 0.598 \\
B$_{11}$C$^{e(11, 12)}$+(C-B-C) & 0.793 \\
 & $\thicksim$0.74 \footref{note:2} \\
\hline 
A chain-centred site & \\
B$_{12}$+(C-C-C) & 1.445 \\
 & $\thicksim$1.69 \footref{note:2} \\
\hline
Disordered chain & \\
B$_{11}$C$^{p(1)}$+(B-C-C) & 2.521 \\
B$_{11}$C$^{p(1)}$+(C-C-B) & 2.598 \\
\hline
Distorted Bipolar Defect & \\
\#1 & 0.628 \\
\#2 & 1.066 \\
\#3 & 1.152 \\
\hline
\hline
\end{tabular}
\end{minipage}
\label{tab:1}
\end{table}
\section{III. Superatom-Special quasirandom structure technique}
	\indent{Modelling of configurational disorder beyond dilute defects in complex crystal structures, such as B$_{4}$C, using the complete mathematical apparatus of Sanchez, Ducastelle, and Gratias~\cite{Sanchez1984} is often impractical. Furthermore, it is not necessary to take into account all kinds of defects in disorder models, while many of them might be neglected due to high total energies. Instead, in such cases, one can try to reveal the important physics by focusing on the modeling of low energy disordered patterns. Here, we suggest that configurationally disordered B$_{4}$C can be viewed as a disordered distribution of different superatoms, each defined based on knowledge of low energy dilute defects. We suggest the superatom-special quasirandom structure (SA-SQS) technique to model this situation in B$_{4}$C. In this approach, the configuration of the superatoms is modeling a random alloy pattern according to the special quasi-random structure (SQS) approach. In our case, a 12-atom icosahedron together with a 3-atom chain were treated as a superatom unit. The first intuitive way to define such superatom basis is to identify the 15 atoms of the B$_{4}$C unit cell as the superatom basis, see Fig.~\ref{fig:4_1}. However, to allow for the description of the two types of dilute defects that we find to have distinct lowest formation energy, i.e. the bipolar defect and the polar-up carbon rotation, a second way of defining the superatom basis was introduced, see Fig.~\ref{fig:4_2}. It replaces, within the basis, the polar-down sites from a single icosahedron with the corresponding polar down sites of neighboring icosahedra with bonds to the original polar-up sites. A configurational disorder, based on either the bipolar defect or the carbon rotation among the polar sites, can then be created.}\\
	\indent{For the creation of disordered configurations of superatoms, we used equal concentration of each included superatom type and the SQS technique to randomly distribute them with an aim to get Warren-Cowley short-range-order parameters ($\alpha$$_{i}$, $i$=1-4) for the first four coordination shells to be as close to zero as possible. As mentioned in the previous paragraph, we introduced two kinds of superatom bases. Both bases allow for six different superatom types, respecting to the position of the icosahedral carbon atom at any of the six polar sites (see Fig.~\ref{fig:4}).
	
	In the present work, six different disordered configurations, constructed using SA-SQS technique, were considered: We include configurations with two types of superatoms (modeling bipolar defects), three types of superatoms (modeling rotation of carbon atoms on the three polar up sites), four types of  superatoms (modeling rotation on polar up sites in combination with one type of bipolar defect), and six types of superatoms (modeling rotation of carbon atoms on all six polar sites) The structures used in the calculations were based on 3x3x3 (405 atoms) and 4x3x3 (540 atoms) supercells. For each disordered configuration, atomic coordinates and the cell shape were optimized in a series of different fixed-volume calculations.}\\
	\indent{The Gibbs free energy with a zero pressure condition, were calculated for all configurations, where the configurational entropy was determined within the mean-field approximation. The transition temperatures were then estimated from a plot of the free energy as a function of the temperature \emph{T}. The equation for the free energy calculation is given by;}\\
	\begin{equation}
		G^{\gamma}=E^{\gamma}-TS^{\gamma},
		\label{eq:2}
	\end{equation}
where \emph{G}$^{\gamma}$, \emph{E}$^{\gamma}$ and \emph{S}$^{\gamma}$ stand for the Gibbs free energy, the total energy and the configurational entropy of the system with configuration $\gamma$. The entropy \emph{S} was obtained within a thermodynamics limit, i.e. the number of particles approaches infinity, using Stirling's approximation of the binomial distribution. A general expression for the configurational entropy \emph{S} in this so-called mean-field approximation is given by;\\
	\begin{equation}
		S=-k_{B}N\sum_{i=1}^{n}x_{i}ln(x_{i}),
		\label{eq:3}
	\end{equation}
where \emph{N} and \emph{n} are defined as the number of superatom sites in the supercell and the number of superatom types included in the supercell, respectively. \emph{x$_{i}$} refers to the concentration of type-\emph{i} superatom. If the concentration for each type of superatom is assigned to be equal, the entropy per superatom site is thus reduced to;\\
	\begin{equation}
		S=k_{B}ln(n),
		\label{eq:4}
	\end{equation}
where \emph{n} is the number of types of superatoms defined for each configuration.\\
\indent{To investigate the effects of pressure on the configurational stability of B$_{4}$C, the total energies at different fixed volumes of some selected configurations were fitted to the 3$^{rd}$ order Birch-Murnaghan equation of state (EOS)~\cite{Murna1944, Birch1947}. The Gibbs free energy of a zero pressure condition according to Eq.~\ref{eq:2} is, therefore, modified and given by;}\\
	\begin{equation}
		G^{\gamma}=E^{\gamma}+PV^{\gamma}-TS^{\gamma},
		\label{eq:5}
	\end{equation}
where \emph{P}, and \emph{V}$^{\gamma}$ are defined as an applied pressure, and the equilibrium volume of the system of configuration $\gamma$ corresponding to the pressure \emph{P}, respectively.
\section{IV. Results and discussion}
\subsection{a. Dilute defects}
\indent{Based on our calculations, we find that the different sites of the unit cell of ideally ordered B$_{4}$C are non-equivalent in defect formation aspect to a higher degree than what has been previously realized. The calculated defect formation energies per defect of dilute configurational defects according to Eq.~\ref{eq:1} are presented in TABLE~\ref{tab:1}. Most of them are calculated in a supercell of 120 atoms (2x2x2 repetition of the unit cell). All defects increase the energy with respect to the ideally ordered B$_{4}$C. However, these defective structures are still very stable with respect to $\alpha$-boron and diamond. This is not surprising based on their dilute character and the high stability of the perfect B$_{4}$C phase. Note that the atomic position of the carbon atom in the icosahedron is denoted with \emph{p} or \emph{e} depending on its position at the polar and equatorial sites respectively, and the numbers refer to the positions according to the notation in Fig.~\ref{fig:2}. For example, the notation \emph{p(1)} corresponds to an icosahedron with a carbon atom positioned according to Fig.~\ref{fig:2}, which is the case for all icosahedra if the ground state (GS)-configuration, B$_{11}$C$^{p(1)}$+(C-B-C), is considered. The defect with the lowest energy with respect to the (GS)-configuration, is the bipolar defect (\emph{$\Delta$E$_{defect}$} = 0.229 eV), illustrated in Fig.~\ref{fig:3}. As shown in TABLE~\ref{tab:1}, our calculated energy is in good agreement with the result calculated by Raucoules \emph{et al.}~\cite{Vast2011} where the supercell's sizes used for the calculations are the same 2x2x2. On the other hand, the bipolar defect formation energy provided by Widom and Huhn~\cite{Widom2012} is almost 2 times larger compared to our result. We suggest that the discrepancy in the formation energy is involved with the difference in the supercell sizes. Since their calculations were done in a 3x1x1 supercell, it is questionable that the supercell is large enough to avoid uncontrolled defect-defect interaction effects. To verify an accuracy in our result, we perform a calculation of the bipolar defect formation energy in a 3x3x3 supercell. The result shows that the formation energy in the 3x3x3 supercell differs only by 0.03 eV/defect from that of the 2x2x2 case. Furthermore, these bipolar defects have been reported to be presented in the structure, as native defects formed during the synthesizing process~\cite{Vast2011}.}\\
\indent{Another type of defect, coming after the bipolar one, is a rotation of the icosahedral carbon atom among the polar-up sites (\emph{$\Delta$E$_{defect}$} = 0.232 eV), i.e., the substitution of the carbon atom at \emph{p(2, or 3)} position instead of \emph{p(1)} position. Note that moving the icosahedral carbon atom to either \emph{p(2)} or \emph{p(3)} are equivalent and they thus give the same formation energies. This type of defect has its formation energy only slightly larger than that of the bipolar defect. However, when even larger, 3x3x3 supercells are used, the formation energies of bipolar and polar-up carbon rotation defects are actually changing order even though the defect formation energies change with about 0.04 eV/defect, meaning that these two types of defects are practically equal in energy. Vast \emph{et al.}~\cite{Vast2009} suggested that it is the random substitution of carbon atoms among all the six polar sites of the icosahedra that results in the rhombohedral symmetry (\emph{R$\bar{3}$m}) of B$_{4}$C observed in experiments. A similar conclusion was made in the works of Widom and Huhn~\cite{Widom2012}. Based on our calculations, the energy it costs to swap a carbon atom to polar-down sites is larger than the swapping of a carbon atom only among the polar-up sites. That is, the energy to flip the icosahedral carbon atom to \emph{p(5, or 6)} and \emph{p(4)} positions, becomes approximately twice (0.46 eV) and 6 times (1.21 eV), respectively of that of the polar-up carbon rotation. The latter case can be thought of as flipping upside-down the icosahedron, thus resulting in the inter-icosahedral C-C bond. Due to its high formation energy, the C-C bond is considered as an unfavorable type of bonding for the B$_{4}$C system as demonstrated also by the case of [(B$_{12}$)+(C-C-C)], i.e. the chain consisting only of carbon atoms, with the formation energy of 1.44 eV. Hence, we expect the icosahedral carbon atoms are more favorable to substitute polar sites in only one side rather than on both up and down sides. We come back to the consequence of polar carbon rotation-based disordered configurations, in section IV:b. Note that apart from those two types of a dilute defect, i.e. the bipolar defect and the polar-up carbon rotation, the remainder provides significantly higher formation energies and thus they are expected to presented in the structure in much lower concentrations.}\\ 
\begin{figure*}[ht!]
     \begin{center}
        \subfigure[]{%
            \label{fig:6_1}
            \includegraphics[width=0.415\textwidth]{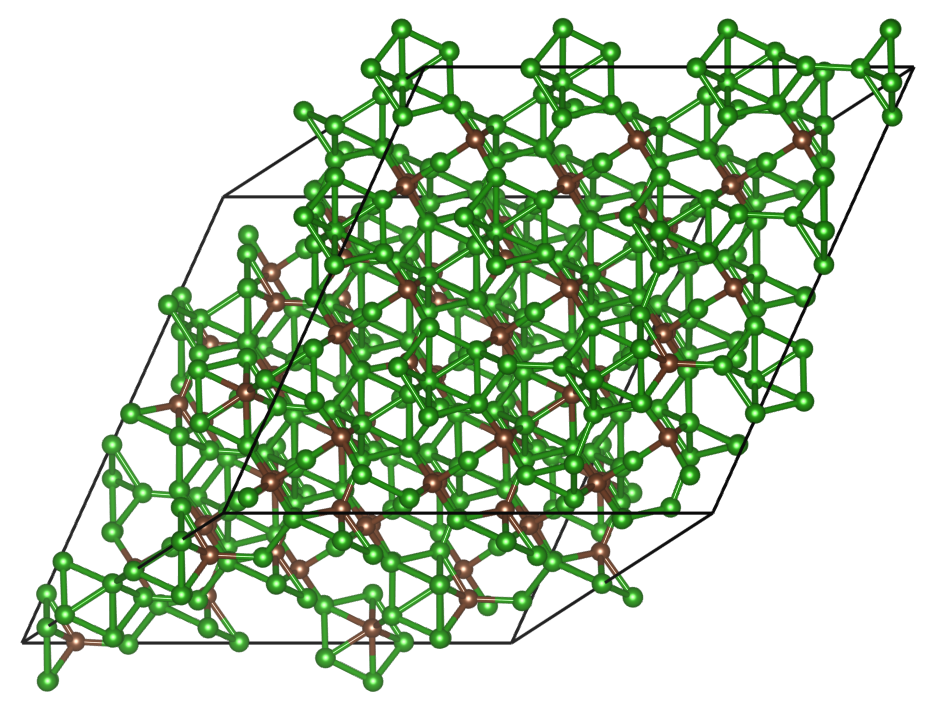}
        }%
        \subfigure[]{%
           \label{fig:6_2}
          \includegraphics[width=0.45\textwidth]{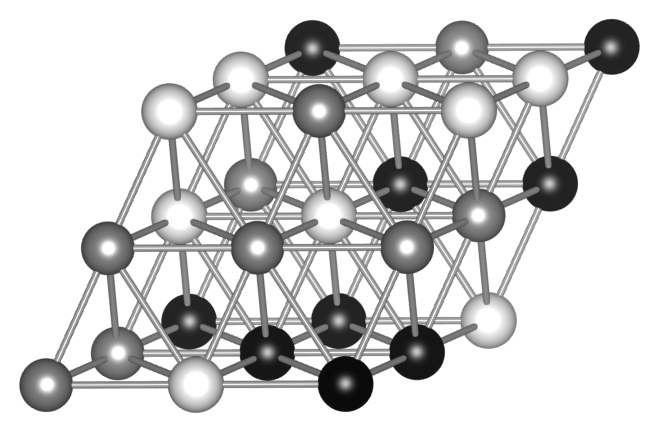}
        }\\ 

    \end{center}
    \caption{%
      (Color online) 3x3x3 super cells of the (3PU)-disordered configuration in B$_{4}$C represented in (a) normal-atom's picture, and (b) super-atom's picture. Green and brown spheres, in (a), represent boron and carbon atoms, respectively. Black, white, and gray spheres, in (b), represent the super-atom's type-\emph{A}, \emph{B}, and \emph{C} for Basis-1, respectively.
     }%
   \label{fig:6}
\end{figure*}
\indent{The non-equivalence of sites also goes for the equatorial and the chain sites. Obviously, the three equatorial-up sites \emph{e(7, 8, 9)} are not equivalent to the equatorial-down ones \emph{e(10, 11, 12)}, similar to the case of the polar sites. More importantly, one site of both the equatorial-up and -down sites is different from the other two sites as revealed by their defect formation energies. The 2 equatorial-up sites \emph{e(8, 9)} are different from the other one \emph{e(7)}. Also, the 2 equatorial-down sites \emph{e(11, 12)} are not equivalent to the other equatorial-down site \emph{e(10)}. Especially for the equatorial down sites, the difference in the formation energy between those non-equivalent sites is as much as 0.2 eV, which is comparable to the formation energy of the bipolar defect as well as of rotating the carbon atom among the polar-up sites. In the case of the chain sites, swapping between boron and carbon atoms sitting on the chain sites in the opposite direction, i.e. [(B$_{11}$C$^{p}$)+(B-C-C)] and [(B$_{11}$C$^{p}$)+(C-C-B)], as well results in different formation energies indicating the non-equivalence in the chain sites. To our knowledge, this degree of non-equivalences of the sites in ideally ordered B$_4$C, both in the icosahedron and in the chain, have so far never been pointed out. In summary, based on the energetics of the dilute defects one can assume that the bipolar defect and polar up rotation of icosahedral carbon atoms, can be expected to be present in B$_{4}$C at high concentrations in equilibrium high-temperature conditions. The other types of defects that are at least twice as high in energy can still be present as dilute defects or as consequences of out-of-equilibrium synthesis with, e.g., insufficient diffusion during sample preparation.}
\begin{table*}
\caption{Configurational details of seven disordered configurations in B$_{4}$C. In the first column, (GS) stand for the ground state configuration of B$_{4}$C with a monoclinic distortion, which is a referential configuration. (BD) refers to the bipolar defect-based disordered configuration, meanwhile (\emph{x}P(U, or D)) stands for the disordered configuration constructed based on substituting carbon atoms at the polar(-up, or -down) sites, where \emph{x} is the number of the sites that are substituted by carbon atoms. (RNG) refers to the disordered configuration obtained by randomly distributing all icosahedral carbon atoms among the polar sites in the structure using a random number generator. $\Delta$E, in the second column, denotes the difference in energy per superatom (s.a.) of the disordered configurations with respect to the ground state. $\Delta$E$^{form}$, in the third column, denotes the formation energy with respect to a phase decomposition into $\alpha$-boron and diamond.}
\begin{tabular}{l @{\hspace{0.5cm}} c @{\hspace{0.5cm}} c @{\hspace{0.5cm}} l @{\hspace{0.5cm}} c @{\hspace{0.5cm}} c @{\hspace{0.5cm}} c}
\hline
\hline
Configuration & $\Delta$E & $\Delta$E$^{form}$ & Types of & Supercell size & Lattice parameters & angles\\
& (eV/s.a.) & (eV/atom) & superatom & (Number of atoms) & \emph{a}, \emph{b}, \emph{c} (Å/unitcell) & $\alpha$, $\beta$, $\gamma$ ($^{\circ}$)\\
\hline
GS & 0 & -0.130 & \emph{A} & 4x3x3 (540) & 5.209, 5.209, 5.059 & 66.01, 66.01, 65.14\\
GS \footnote{{Ref.~\cite{VI2009}} - (LDA-PP) \label{note:4}} & - & - & - & 1x1x1 (15) & 5.143, 5.143, 5.010 & 66.39, 66.39, 65.57 \\
BD & 0.066 & -0.126 & \emph{A'}, \emph{D'} & 4x3x3 (540) & 5.209, 5.209, 5.060 & 66.01, 66.00, 65.14\\
3PU & 0.083 & -0.125 & \emph{A}, \emph{B} \emph{C} & 3x3x3 (405) & 5.159, 5.159, 5.159 & 65.73, 65.73, 65.73\\
3PU+1PD & 0.179 & -0.118 & \emph{A}, \emph{B}, \emph{C}, & 4x3x3 (540) & 5.173, 5.173, 5.136 & 65.79, 65.79, 65.58\\
 & & & \emph{D} & & & \\
6P & 0.226 & -0.115& \emph{A}, \emph{B}, \emph{C}, & 4x3x3 (540) & 5.161, 5.161, 5.161 & 65.72, 65.72, 65.72\\
 & & & \emph{D}, \emph{E}, \emph{F} & & & \\
BD+3PU & 0.163 & -0.119 & \emph{A'}, \emph{B'}, \emph{C'}, & 4x3x3 (540) & 5.172, 5.172, 5.135 & 65.80, 65.79, 65.58\\
 & & & \emph{D'} & & & \\
BD+6P & 0.222 & -0.116 & \emph{A'}, \emph{B'}, \emph{C'}, & 4x3x3 (540) & 5.161, 5.161, 5.161 & 65.73, 65.73, 65.73\\
 & & & \emph{D'}, \emph{E'}, \emph{F'} & & & \\
RNG & 0.553 & -0.093 & - & 3x3x3 (405) & 5.151, 5.162, 5.172 & 65.66, 65.76, 65.78\\
Experiment \footnote{Ref.~\cite{Morosin1995} - Neutron diffraction \label{note:5}} & - & - & - & - & 5.163, 5.163, 5.163 & 65.73, 65.73, 65.73 \\
Experiment \footnote{Ref.~\cite{Morosin1987} - X-ray diffraction \label{note:6}} & - & - & - & - & 5.155, 5.155, 5.155 & 65.68, 65.68, 65.68 \\
\hline
\hline
\end{tabular}
\label{tab:2}
\end{table*}
\subsection{b. Disorder configurations of B$_{4}$C}
\indent{According to what was indicated in the previous subsection, we conclude that two types of defects, (1) the bipolar defect and (2) a rotation of the icosahedral carbon atom among the polar-up sites, are the most probable defects to be present in the structure with high concentrations at equilibrium. The disordered configurations were, therefore, constructed mainly based on such two types of dilute defects. However, as the disorder of carbon atoms on all six polar sites has beens discussed in the literature, also this type of disorder was considered. As described in section III, we suggest two kinds of superatom bases for constructing the disordered configurations, as shown in FIG.~\ref{fig:4}. Clearly for Basis-1 superatom, a whole icosahedral structure with the 3-atom chain is defined as one superatom. By using this basis, the disordered configuration, in which the icosahedral carbon atom in each icosahedron can substitute any of the six polar sites or any of only the three polar-up sites, can be obtained. However, Basis-1 alone is not enough to describe the bipolar defects. This is because the bipolar defect is not only the coexistence of B$_{12}$ and B$_{10}$C$_{2}$ icosahedra, but their particular arrangements are also needed to be taken into consideration, as can be seen in the energy difference between the bipolar defect and the distorted bipolar defects. To handle such difficulties, a use of Basis-2 superatoms becomes important. 

In the present work, seven different disordered configurations have been considered. Among these seven configurations, there are six, constructed using our SA-SQS technique (see FIG.~\ref{fig:6} for the SA-SQS describing polar up rotation), whilst the seventh configuration was created using a random number generator (RNG) to distribute all of the 27 icosahedral carbon atoms among the 162 polar sites in the supercell in an uncorrelated manner. The configurational details of seven different disordered configurations are given in TABLE~\ref{tab:2}. The short-range-order parameters $\alpha$$_{i}$ are zero up to the fourth shell for the bipolar defect-based disordered configuration. For the 3PU-configuration, where the icosahedral carbon atoms are allowed to substitute only the three polar-up sites, the short-range-order parameters are zero for the first and second shells. Small non-zero SRO values are present for the other multicomponent cases and supercell sizes larger than 4x3x3 would be required to obtain an ideal random alloy model. Unfortunately, due to the 15 atoms per superatom basis, even larger supercells would not be computationally affordable in the present work. As a consequence, we have limited the supercell size to 4x3x3 superatoms, corresponding to 540 individual atoms.

In Table~\ref{tab:2} the energy with respect to the configurational ground state, the type of superatoms included, the supercell size, and the lattice parameters, for each case are reported. 
The disordered configuration based on bipolar defects (BD) is found to be 0.066 eV per superatom higher in energy than the configurational ground state. The rotation of carbon atoms among polar up sites (3PU) is 0.083 eV per superatom higher than the ground state. Then follows in order of increasing energy the combination of bipolar defects with polar up rotation in Basis-2, 3 polar up rotation together with one of the polar down sites, and the two cases with 6 types of superatoms with carbon disordered among all polar sites in the two respective bases. Highest in energy is the configuration created with random number generator distribution of polar site carbon. Again the phase stability against phase decomposition into $\alpha$-boron and diamond for all disordered configurations given in TABLE~\ref{tab:2} are considered. Calculating the formation energy indicates that they are stable against the phase decomposition, where their formation energies fall into the range between -0.130 eV/atom and -0.093 eV/atom for the most stable (GS)-configuration and for the least stable (RNG)-configuration, respectively.

By comparing the lattice parameters between the disordered configurations and the ground state, effects of disorder on the symmetry of the system can be found. The substitutional disorder of carbon atoms among all the six polar sites in the (6P)-configuration fully restores the rhombohedral symmetry (\emph{R$\bar{3}$m}) on average. This is in line with what has been suggested~\cite{Vast2009, Widom2012}. However, here we find that the substitutional disorder of carbon atoms among the three polar-up sites only is sufficient to restore the rhombohedral system. It is, however, lacking inversion symmetry. This loss of the inversion symmetry in the (3PU)-configuration thus yields the space group \emph{R$3$m, \#160}. In addition to a rotation of the carbon atom among three polar-up sites, we have, as well, considered the (3PU+1PD)-configuration, which allows the icosahedral carbon atoms to substitute the three polar-up and one of the polar-down sites. This configuration is less favorable compared to the (3PU)-configuration because \emph{(1)} the energy required to move a carbon atom  to the polar-down sites is rather high, as indicated in section II, and \emph{(2)} there is the possibility for intericosahedral C-C bonds to be formed. The degree of monoclinic distortion, in the case of (3PU+1PD)-configuration, is smaller, placing its structure somewhere in between those of the (GS)- and the (3PU, or 6P)-configurations. We suggest that the decrease in monoclinic distortion in this case is due to the competitive effects between an equal distribution of the carbon atoms among the three polar-up sites that is trying to restore the rhombohedral symmetry and a substitution of the carbon atoms at only one of the polar-down sites that induces the monoclinic distortion. The existence of bipolar defects does not induce any change with respect to the the lattice parameters and the angles. They are all practically the same as of the (GS)-configuration. The monoclinic distortion still remains in the (BD)-configuration.

The a, b, and c lattice parameters ($\alpha$, $\beta$, $\gamma$ angles) are in fact not exactly identical even for the 6P-, the 3PU-, and the BD+6P-configurations.  However, the magnitude of the difference is small (of the order of 0.001 \AA (degree)). However, from a practical point of view they are the same. Thus, practically, and exactly in a hypothetical infinite size supercell, the rhombohedral symmetry (\emph{R$\bar{3}$m}) is restored on average for the (6P)-, and the (BD+6P)-configurations, whilst the lower symmetry (\emph{R$3$m}) without the inversion is restored for the (3PU)-configuration. The lattice parameters and the angles of those having the rhombohedral symmetry (both \emph{R$3$m} and \emph{R$\bar{3}$m}) are in very good agreement with the experimental result obtained from the neutron diffraction~\cite{Morosin1995} as given in TABLE~\ref{tab:2}.\\
\indent{The configurational results reveal the monoclinic distortion decreases in the (BD+3PU)-configuration, which is more or less the same to that of the (3PU+1PD)-configuration. On the other hand, the (BD+6P)-configurations, as mentioned above, has the rhombohedral symmetry (\emph{R$\bar{3}$m}). The explanation for both the smaller monoclinic distortion in the (BD+3PU)-configuration and the restoration of the higher symmetry in the (BD+6P)-configuration can be given similarly as in the case of the (3PU+1PD)- and (6P)-configurations, respectively. The (RNG)-configuration is the most configurationally disordered, since the icosahedral carbon atoms are allowed to sit non-specifically on any of the polar sites in the structure, limited only by the supercell size. As a consequence, the number of carbon atoms contained in each icosahedron, in principle, can be varied from zero to six. In our case, the number of carbon atoms in each icosahedron is varied from zero to two. To specify the symmetry of the (RNG)-configuration, we assume an infinitely large B$_{4}$C system. The probability that all polar sites are occupied by carbon atoms becomes equal since, in the (RNG)-configuration, the icosahedral carbon atoms randomly substitute any polar sites in the icosahedra. Its symmetry should eventually be similar to that of the (6P)-configuration, i.e. the space group \emph{R$\bar{3}$m}. This (RNG)-configuration is the energetically most unfavorable among all of our disordered configurations. The reason is that the possibility of having unfavorable  carbon-carbon bonds in the structure is very high.}

\begin{figure}[h]
                \centering
                \includegraphics[width=0.7\linewidth, angle=270]{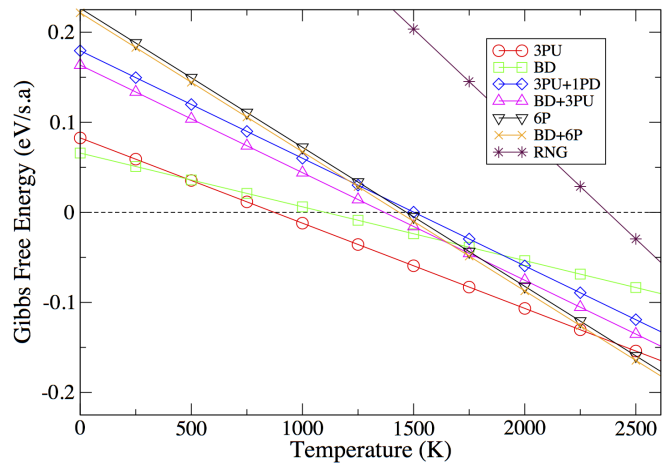}
                \caption{(Color online) Difference in Gibbs free energy (\emph{P}=0) for seven disordered-configurations, relative to the ground state configuration (dashed line), as a function of temperature.}
                \label{fig:7}
        \end{figure}

FIG.~\ref{fig:7} represents the Gibbs free energy at a zero pressure condition, plotted as a function of the temperature and with the free energy of the ground state configuration taken as the zero value.  The entropy \emph{S} and the free energy \emph{G} were calculated from the formula given in section III. It should be noted that the ordered ground state has zero entropy.

Even though the (BD)-configuration has the lowest energy of the disordered configurations, the transition from the (GS)- to the (BD)-configuration is not predicted to happen. Instead we observe a transition from the (GS)- to the (3PU)-configuration at the temperature 870 K. The reason is that the (3PU)-configuration with three types of superatoms has a larger configurational entropy than the (BD)-configuration with only two types. At this temperature, the icosahedral carbon atoms should gain enough energy and equally distribute themselves among the polar-up sites. Keeping in mind the findings for lattice symmetry above, we thus predict that there should be a transition of B$_{4}$C from the (GS)-configuration, where the monoclinic distortion is existing, to the rhombohedral symmetry (\emph{R$3$m}) at temperatures above 870 K under equilibrium conditions.

In our simulation, we observe a second configurational phase transition at 2325 K from the (3PU)- to the (BD+6P)-configuration, where carbon atoms are distributed over all polar sites modeled with the Basis-2 type superatoms. At this transition temperature, the higher rhombohedral symmetry (\emph{R$\bar{3}$m}), in which the inversion symmetry is applied on average, is obtained. It is worth noting that as the second transition takes place at a very high temperature which is close to the melting point of B$_{4}$C ($\thicksim$2650 K), the vibrational effects clearly cannot be neglected, i.e. the vibrational contributions both of the energy and the entropy should be taken into account to give a good description of the stability of phases at such high temperature.

We do not observe any phase transition related to the types of disorder that involves 4 types of superatoms described by either Basis-1 or -2. As expected from the (RNG)-configuration, the contribution of the configurational entropy is much higher than those of the other six configurations, as can be seen from its high slope in the Gibbs free energy diagram. The expression of the configurational entropy per superatom \emph{S}$^{RNG}$, obtained by using the Stirling's approximation, is given by;\\
	\begin{equation}
		S^{RNG}=-6k_{B}[\frac{1}{6}ln(\frac{1}{6})+\frac{5}{6}ln(\frac{5}{6})].
	\end{equation}
Even though the entropy of the (RNG)-configuration is large, its free energy is still distinctly high and it is not even stable at high temperature. This is because there are many unfavorable intericosahedral C-C bonds existing in the supercell. We, consequently, disregard the possibility to have it as the high temperature phase.

Our findings are different from what predicted by Widom and Huhn~\cite{Widom2012, Widom2013}. Based on their study, the (GS)-configuration is stable up to 600 K before it undergoes the configurational phase transition to the (6P)-configuration, where they claimed that carbon atoms became disordered among all six polar sites in the icosahedra, thus yielding the space group \emph{R$\bar{3}$m} on average. Their model were based on partition function creation with input from a limited number of individual configurations in small supercells. In such an analysis there exists inaccuracies in defect energetics due to defect-defect interactions and incomplete sampling of random-like configurations. On the other hand, our results reveal two configurational phase transitions, where the configuration with the full rhombohedral symmetry (\emph{R$\bar{3}$m}) is stable at much higher temperature. Note that our calculations are based on the mean-field approximation, which is known to overestimate ordering transition temperatures by approximately a few hundreds Kelvin. \\ 
\indent{Based on our results we suggest an explanation to the experimental observation of rhombohedral symmetry (\emph{R$\bar{3}$m}) without monoclinic distortion even at low temperatures: The atomic diffusion in B$_{4}$C is quenched above the critical ordering temperature, i.e. higher than 870 K predicted as the first transition temperature, probably because of very strong covalent bonds in boron-carbon system, thus preventing the (GS)-configuration with the monoclinic distortion from appearing. Experimentally, a crystalline boron carbide is normally obtained when synthesized or annealed from initially amorphous phase at very high temperature, where the atomic diffusion is active, typically in the range between 1600 and 2000 K~\cite{Camille2013, Sonber2013}. In such a range of temperature, the crystalline phase formation temperature can possibly fall into either \emph{(1)} the region where the (BD+6P)-configuration is stable or \emph{(2)} the region where the (3PU)-configuration is stable. In the first case, the full rhombohedral symmetry (\emph{R$\bar{3}$m}) will obviously be achieved on average corresponding to what has been reported in literature. Meanwhile, in the second case, the predicted rhombohedral symmetry will be lacking inversion symmetry thus being specified by \emph{R${3}$m} on average. A lack of inversion symmetry is in contrast to the reports of \emph{R$\bar{3}$m} symmetry in the experimental literature. One possibility is that real existing boron carbide, although possessing \emph{R$\bar{3}$m} symmetry globally, could display a short-range order tendency towards either predominantly polar-up or predominantly polar-down disorder, co-existing in different domains. In such a case the global symmetry would be \emph{R$\bar{3}$m} with inversion symmetry present on average, while the energetics and the number of boron-boron, carbon-carbon, and boron-carbon bonds of the system would still be better described by our (3PU)-configuration for polar disorder on one pole only. These questions require theoretical methods going beyond the mean-field description of configurational disorder presented here.}\\
\indent{Also, future experimental work may shed light on the local configurations of polar carbon in B$_{4}$C. However, we note that the difference in atomic form factor for X-ray diffraction between boron and carbon is very small and the lattice parameters, as seen in TABLE~\ref{tab:2}, of the (3PU)- and the (BD+6P)-configurations are practically identical. This similarity in X-ray form factors, the almost identical lattice parameters, and the fact that the difference between such \emph{R$\bar{3}$m} and \emph{R${3}$m} configurations discussed above only involves partial occupation numbers of one out of three carbon atoms in the structure, should cause considerable challenges for experimental determination of this aspect.}
\subsection{c. Pressure effects}
Our approach makes it possible to also investigate the impact of high pressure on the configurational stability of the disordered B$_{4}$C phases. In the present work, two different configurations are considered, i.e. the (GS)- and the (3PU)-configuration. As mentioned in section III, the total energies at different fixed volumes of these two configuration are fitted to the 3$^{rd}$ order Birch-Murnaghan equation of state (EOS). The effects of pressure on the configurational stability are then determined by the Gibbs free energy given in Eq.~\ref{eq:5}. We have found that the applied pressure has a very small effect on the configurational stability of B$_{4}$C. The rate of change of the free energy is on the order of 10$^{-5}$ eV\textperiodcentered{GPa$^{-1}$}\textperiodcentered{atom$^{-1}$}. In fact, this is because of the very high value of bulk modulus of B$_{4}$C. The bulk moduli obtained for the (GS)- and the (3PU)-configurations are 237.6 and 237 GPa, respectively, which are in good agreement with the experimental value, i.e. 247 GPa, given by Gieske \emph{et al.}~\cite{Gieske1991}. We also observed that the transition temperature of B$_{4}$C from the (GS)- to the (3PU)-configurations is lifted up with pressure by the rate of 6 K per GPa. This is due to the slightly smaller volume of the (GS)-configuration as compared to the (3PU)-configuration.

Based on our calculations, by applying a pressure of 15 GPa, the transition temperature would be raised by 90 K, approximately. This increase in ordering temperature with pressure might be tried as a route to reach the predicted ordered (GS)-configuration experimentally. However, recently  Mikhaylushkin \emph{et al.}~\cite{Zunger2013} predicted that ordered B$_{4}$C becomes unstable with respect to phase separation into diamond and pure boron at pressures above 20 GPa. This has the consequence that the use of even higher pressure in attempts to stabilize the (GS) monoclinic distorted phase in a temperature range where diffusion is active becomes limited by appearance of phase separation.

However, it should be noted that Fujii \emph{et al.}'s experiments~\cite{Fujii2010} demonstrated that B$_{4}$C is stable at pressures above 120 GPa, probably due to low diffusion at the low temperatures used, which made phase separation impossible.

\subsection{d. Electronic density of states}
In this subsection, the band gap and electronic density of states (DOS) of the configurationally disordered B$_{4}$C are discussed. Confirmed by experiments~\cite{Werheit1971, Werheit1991, Werheit1997, Werheit2006}, boron carbide is a semiconducting material with a wide range of energy gaps, depending on its stoichiometry and quality. The indirect band gaps of B$_{4}$C, having been reported so far, fall into the range between 0.48 and 2.5 eV~\cite{Werheit1971, Werheit1991, Werheit1997, Werheit2006}. However, larger band gaps, within a range from 2.6 up to 3.0 eV, are obtained from LDA-based theoretical calculations~\cite{Bylander1990, Shirai_12010, VI2009}. This issue has, hence, been questionable because it is known from other materials that both LDA- or the GGA-based electronic structure calculation typically underestimates, not overestimates, band gaps. Based on our calculations, the band gap of as large as 3 eV, with the use of GGA-PBE for the exchange-correlation functional, is obtained for the (GS)-configuration. A better accuracy for band gap calculations is expected to be achieved by using the hybrid functional (HSE) for the exchange-correlation functional. It is known to resolve the problem of underestimating band gaps in standard LDA- and the GGA-based calculations. The reason is that, within the HSE, the non-local Hartree-Fock exchange energy is included. Here we use this more reliable method to investigate the band gap of ideally ordered (GS)-B$_4$C. As expectedly, our results show that the band gap of the (GS)-configuration becomes even wider, $\thicksim$4 eV, when the HSE06 functional, is used. However, calculating the band gaps from those disordered configurations constructed within very large supercells using the HSE06 is not doable due to its extremely expensive computational cost. Therefore, instead of using HSE06, we employ the modified Becke-Johnson (MBJ) exchange potential, which yields band gaps with an accuracy on about the same level to that obtained from the hybrid functional, but much less computational effort. Within the MBJ exchange potential in combination with the GGA-PBE correlation, the band gap of the (GS)-B$_{4}$C is $\thicksim$3.7 eV. We thus prove, with these more reliable methods, that the huge difference, between the computed and the experimentally measured bandgaps for B$_{4}$C does not arise from inaccuracies in the used exchange-correlation functionals. Consequently, we hypothesize that a shrinkage of the band gap in B$_{4}$C originates from configurational disorder, either the equilibrium types discussed above, or owing to defects caused by out of equilibrium synthesis. 
\begin{table}
\caption{Electronic band gap of the configurationally disordered B$_{4}$C. The abbreviations, used in the first column, are the same as given in TABLE~\ref{tab:2}.}
\begin{tabular}{l @{\hspace{0.5cm}} c @{\hspace{0.5cm}} c @{\hspace{0.5cm}} c}
\hline
\hline
\multicolumn{3}{r @{\hspace{0.5cm}}}{Band gap (eV)} \\
\cmidrule(r){2-4}
Configuration & GGA-PBE & MBJ-GGA & HSE06 \\
\hline
GS (ref.) & 3.00 & 3.72 & 4.13 \\
BD & 2.66 & 3.36 & - \\
3PU & 2.98 & 3.72 & - \\
3PU+1PD & 2.87 & 3.60 & - \\
6P & 2.91 & 3.69 & - \\
BD+3PU & 2.63 & 3.31 & - \\
BD+6P & 2.72 & 3.40 & - \\
RNG & 2.01 & 2.76 & - \\
\hline
\hline
\end{tabular}
\label{tab:3}
\end{table}
It is worth noting that, in some senses, the band gap concept can be replaced by mobility gap, when one is dealing with disordered materials, because overlapping between localized states in a band gap due to, e.g. different kinds of defects, and extended stated from the band edges by, e.g. structural disorder, can take place, thus resulting in a smooth transition between these states. This is, somehow beyond our scope, since we do not investigate which states are dominated by defects or disordering effects. Consequently, only the band gap term will be using in the present work and the effects of configurational disorder on the band gap of B$_{4}$C will be discussed in detail in the paragraphs below.

We then investigate how different types of dilute defects affect the band gap of B$_{4}$C based on the GGA-PBE functional. Except the rotation of the icosahedral carbon atom at the polar sites, the other dilute defects created by substituting the carbon atom in the icosahedron, including the distorted bipolar defects, can reduce the band gap with about 0.15-0.3 eV. On the other hand, the defective chain configurations can even more shrink the band gap down to (1) 1.6 eV for [B$_{12}$+(C-C-C)] by creation of the mid-gap states~\cite{Dekura2010, VI2009}, and (2) 2.4 eV for [B$_{11}$C$^{p(1)}$+(C-C-B)] or [B$_{11}$C$^{p(1)}$+(B-C-C)] by causing additional states near the conduction band edge.

\begin{figure}[htb]
		\centering
		\includegraphics[width=\linewidth]{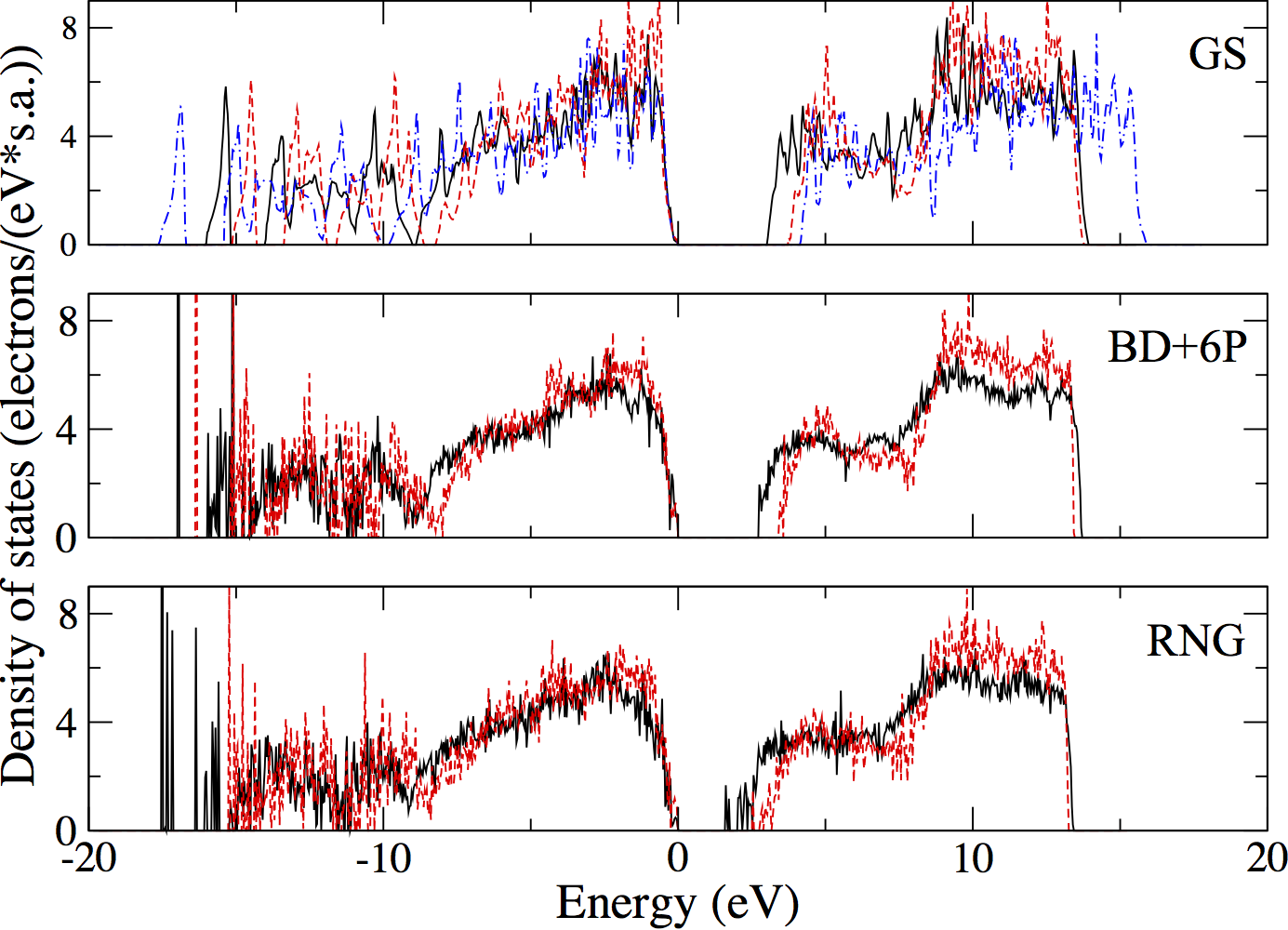}
		\caption{ (Color online) Density of states (DOS) of the ordered/disordered B$_{4}$C. The black solid line, the red dashed line, and the blue dashed-dotted line in the row labeled "GS" indicate the electronic DOS of the (GS)-configuration, obtained by using the GGA-PBE, the MBJ-GGA, and the HSE06, respectively for the exchange-correlation functional. The the second and the third rows represent the electronic DOS of the (BD+6P)- and the (RNG)-configurations, respectively with the GGA-PBB (the solid black line), and the MBJ-GGA (the red dashed line) exchange-correlation functionals.}
		\label{fig:8}
	\end{figure}
\indent{The energy gaps of the disordered configurations, constructed using the SA-SQS technique, are shown in TABLE~\ref{tab:3} and the electronic density of states for the GS, BD+6P, and RNG configurations are shown in Figure 8. As clearly seen in the table, there is a very small change in the band gap ($\thicksim$0.1 eV) of the disordered configurations, induced by the random substitution of the icosahedral carbon atom among the polar sites on each icosahedron, i.e. (3PU)-,(3PU+1PD)-, and (6P)-configurations, whereas a reduction of the band gap by $\thicksim$0.3 eV is found from those bipolar defect (BD)-based disordered configurations. These results are in accord with what is obtained from the dilute defect calculations. The band gap of the (RNG)-configuration is considerably smaller as compared to the (GS)-configuration,  2.01 (2.76) eV for GGA-PBE (MBJ-GGA) functional, due to extra electronic states adjacent to the conduction band edge (see FIG.~\ref{fig:8}). There are also mid-gap states, in the (RNG)-configuration, at 1.6 and 2.5 eV above the valence band edge for the GGA-PBE and MBJ-GGA functionals, respectively. These additional states seems to be similar to the cases of the defective chain, i.e. (C-C-C)-, (B-C-C)-, and (C-C-B)-chains, as mentioned above. We know from previous works~\cite{VI2009, Dekura2010} that the replacement of the central boron atom in the chain by a carbon atom results in mid-gap states. However, in our case, we did not introduce any chain-defect into the configuration. We, therefore, suggest that the additional states, both nearby the conduction band edge and in the middle of the band gap, originate from a high concentration of C-C bonds existing in the (RNG)-structure. The DOS of the (GS)-, (BD+6P)-, and (RNG)-configurations are illustrated in FIG.~\ref{fig:8}. Note that apart from the size of the band gap, the DOS for the (BD)-, (BD+3PU)-, (3PU)-, (3PU+1PD)-, and (6P)-configurations are more or less the same (on average) to that of the (GS)-configuration. Even though our results reveal a trend of band gap decrease with increasing degree of configurational disorder, they do not explain the experimental results quantitatively as one would have expected a smaller bandgap for GGA calculations in that case. Without taking into account those mid-gap states, the (RNG)-configuration, that has the smallest bandgap that is even below (close to) the largest experimental report when the GGA-PBE (MBJ-GGA) functional is used, has a very high value of the free energy, even at high temperature. While we do not expect such a state to appear in equilibrium, its smaller bandgap indicates that the presence of high-energy defects caused by difficulties to reach equilibrium during synthesis might be the cause for the lower than expected experimental bandgap. The large spread in the experimental reports of the bandgap is in itself an indication of the sensitivity of the electronic properties of B$_4$C to structural disorder. This is in accord with the study done by Ivashchenko \emph{et al.}~\cite{VI2009}, in which they investigated the electronic density of states (DOS) of amorphous B$_{4}$C using molecular dynamic simulations. Their results reveal a DOS minimum, instead of a band gap, in the amorphous B$_{4}$C yielding semi-metallic properties.}
\section{V. Conclusions}
\indent{We have studied configurational disorder in B$_{4}$C, using first-principles calculations. A  method to access different disordered configurations, namely the superatom-special quasi random structure (SA-SQS) technique, is demonstrated. The method is a combination of a concept of super-atoms to describe low-energy defects in a complex structure and a distribution of the superatoms like a random alloy using the special quasi-random structure (SQS) technique. By using this SA-SQS technique, the properties of the configurationally disordered B$_{4}$C, e.g. symmetry, and electronic density of states, become accessible. We predict two configurational phase transitions (at P=0 GPa), at $\thicksim$870 K and $\thicksim$2325 K, respectively, from the ordered to increasingly disordered phases, within the mean-field approximation. After undergoing the first transition, the higher rhombohedral symmetry without the inversion (\emph{R${3}$m}) in B$_{4}$C is restored, due to disorder of icosahedral carbon among the three polar up sites in the structure, meanwhile the full rhombohedral symmetry (\emph{R$\bar{3}$m}) reported in experiments is restored after the second transition, where the random substitution of the icosahedral carbon atoms at all six polar sites and the bipolar defects exist in the icosahedral structure of B$_{4}$C with high concentrations. The details of the icosahedral carbon disorder and the difference of the \emph{R${3}$m} and \emph{R$\bar{3}$m} states deserves further attention with methods going again beyond the mean field approximation for configurational thermodynamics, and most importantly, includes vibrational effects.}\\
\indent{Nevertheless, our present level of calculations provides an explanation for why the B$_{4}$C ordered ground state with monoclinic distortion has so far not been observed: The atomic diffusion in B$_{4}$C is likely to be quenched above the critical ordering temperature prohibiting access to the low temperature ordered state. By investigating the impact of high pressure on the configurational stability of the disordered B$_{4}$C phases, we find that the configurationally ordering/disordering transition temperature increases with pressure exerted on B$_{4}$C. Also by considering the electronic density of states, we find that the band gap of B$_{4}$C is sensitive to the degree of disorder, also for configurational disorder in the icosahedra. Such changes in the electronic structure demonstrate that apart from the stoichiometry, also the configuration of both the chain and icosahedra do influence the electronic structure and the band gap of B$_{4}$C.}
 
\section{Acknowledgments}
Financial support by the Swedish Research Council (VR) through the young researcher grant No. 621-2011-4417 is gratefully acknowledged by B.A. The support from the Swedish Research Council (VR) Project No. 2011- 42-59, LiLi-NFM, and the Swedish Government Strategic Research Area Grant in Materials Science to AFM research environment at LiU are acknowledged by S.I.S. Financial support from the KAW project \textquotedblleft{}Isotopic Control for Ultimate Materials Properties\textquotedblright{} is highly appreciated by J.B. The simulations were carried out using supercomputer resources provided by the Swedish national infrastructure for computing (SNIC) performed at the National supercomputer centre (NSC). Carina H\"oglund, Mewlude Imam, and Henrik Pedersen are acknowledged for useful discussions. 


\end{document}